\documentclass[10pt]{article}

\usepackage{latexsym}
\usepackage[namelimits,intlimits,sumlimits]{amsmath}
\usepackage{amsfonts}
\usepackage{amssymb}
\usepackage{epsfig}

\def\N{\mathbb{N}}
\def\R{\mathbb{R}}

\def\leq{\leqslant}
\def\geq{\geqslant}
\def\qed{\hfill $\square$}
\newtheorem{Pro}{Proposition}[section]
\newtheorem{Thm}[Pro]{Theorem}
\newtheorem{Lem}[Pro]{Lemma}

\begin{document}

\title{Route to chaotic
synchronisation in coupled map lattices: Rigorous results}
\date{}
\author{Bastien Fernandez and Pierre Guiraud}
\maketitle

\begin{center}
Centre de Physique Theorique (FRUMAM)\\
CNRS Luminy, Case 907\\
13288 Marseille CEDEX 09, France\\
email: bastien@cpt.univ-mrs.fr and guiraud@cpt.univ-mrs.fr
\end{center}
\bigskip

\begin{abstract}
Two-dimensional mappings obtained by coupling two piecewise increasing
expanding maps are considered. Their dynamics is described when the
coupling parameter increases in the expanding domain. By introducing a
coding and by analysing an admissibility condition, approximations of
the corresponding symbolic systems are obtained. These approximations
imply that the topological entropy is located between two decreasing
step functions of the coupling parameter. The analysis firstly applies
to mappings with piecewise affine local maps which allow explicit
expressions and, in a second step, is extended by continuity to
mappings with piecewise smooth local maps.
\end{abstract}

\section{Introduction}
Coupled map lattices (CML) are discrete models for the nonlinear
dynamics of some extended systems. For instance, they have been used to
model population dynamics, reaction-diffusion systems or alloy
solidification processes. We refer to the article by Kaneko in
\cite{K93} for a review of applications of CML. As dynamical systems
with multidimensional (infinite dimensional) phase space, CML are
also interesting in their own because they often do no satisfy usual
assumptions such as uniform hyperbolicity or existence of Markov partition.

The dynamics of CML is generated by a mapping with two
components. One component is a mapping of the real variable, the local
map, which represents a local forcing at each site of the lattice. The
other component is a convex combination of local states, the coupling
operator, which represents a diffusive interaction between sites. A
parameter, called coupling, measures the strength of the coupling
operator. (The coupling is equal to zero means that the coupling
operator is the identity, i.e.\ that the system is uncoupled.) One of
the main question in the theory of CML is to describe the dynamics
when the coupling increases from zero \cite{B97}. 

For CML with expanding local map, in the domain of couplings where
the mapping keeps expanding, the dynamics has mainly been 
mathematically described for (very) small couplings. In such a domain,
the complete chaotic structure of the uncoupled system persists. That is to
say, there exists an absolutely continuous invariant measure with
decay of correlations \cite{BK97,JP98} and/or the system is topologically
conjugated to the uncoupled system \cite{AF00}. This SRB-measure needs
not be unique if the lattice has an infinite number of sites
\cite{JJ01}.

For stronger coupling but still when the mapping is expanding, effects
of diffusive interaction modify the dynamics. From the
ergodic point of view, phase transitions are expected
\cite{BBCFP01} and were proved to occur in CML with special coupling
operator \cite{GM00}. From the topological point of view, the
complexity of the system (e.g.\ its topological entropy) might 
decrease with coupling but this decay has not been mathematically
proved. One reason is that excepted in special examples \cite{D98}
where a Markov partition and the subsequent transition matrix have
been determined, the symbolic dynamics associated to CML remained unknown. 

When the coupling increases further so that the mapping ceases to be
expanding, chaotic synchronisation takes place \cite{PRK01}. It means that the
diagonal of the phase space attracts (at least) a set of initial
conditions of positive Lebesgue measure \cite{ABS96,KKN92}. 

These results confirm that the dynamics fundamentally differs between
the small coupling domain and the synchronisation region. However,
changes in the dynamics when the coupling increases between these two
domains, the route to synchronisation for short, is not so well
known. The present paper is devoted to the description of these
changes in some CML with two sites. The local maps are chosen
expanding, piecewise increasing and similar to Lorenz maps
\cite{R78}. The route to synchronisation is described in the framework
of symbolic dynamics and translated in terms of topological
entropy. By using a natural coding, upper and lower estimates
of the set of symbolic sequences which imply that the topological
entropy is located between two decreasing step functions of the
coupling. It means that the complexity of the coupled map lattice
becomes smaller and 
approaches the complexity of the synchronised system when the coupling
increases.

The plan of the paper is as follows. We firstly consider
piecewise affine maps. In Section \ref{PAECML}, we obtain an explicit
expression of the admissibility condition of symbolic sequences and we
prove that the symbolic system is conjugated to the CML on its
repeller (Proposition \ref{CONJU}). Moreover, we establish a partial
ordering on symbolic sequences compatible with the ordering of the
coordinates of their images (Proposition \ref{PROFI}). With these
tools provided, we obtain an upper bound and a lower bound for the set
of admissible sequences which both decrease with coupling (Theorem
\ref{DESC}). Computing the corresponding topological entropies gives
upper and lower bounds for entropy of the CML (Figure \ref{BORNE}). In
addition monotonicity of 
the set of admissible sequences in a neighbourhood of the uncoupled
domain is also shown. All these results are given Section 3 and their
proofs are given in Sections 4 and 5. In a second 
step (Section 6), by proving a kind of $C^1$-structural stability
for piecewise increasing expanding CML, the previous bounds are
extended to CML with smooth local maps. Precisely, Theorem
\ref{LAST} shows that small perturbations of the local map implies
small perturbations of the complete dynamical picture in the coupling
domain.

\section{Piecewise affine expanding CML and their coding}\label{PAECML}
\begin{figure}
\epsfxsize=5truecm 
\centerline{\epsfbox{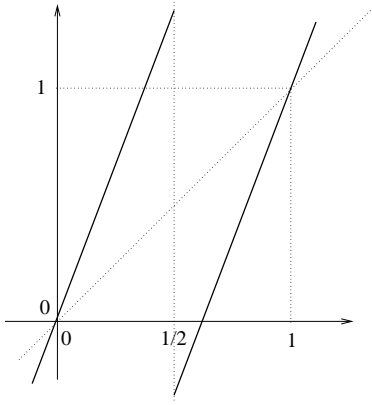}}
\caption{The local map $f$.}
\label{LOCAL_FIG}
\end{figure}

Coupled map lattices of two sites are discrete dynamical systems in
$\R^2$ generated by the one-parameter family of maps
$F_\epsilon:\R^2\rightarrow\R^2$ defined by 
\begin{equation}
F_\epsilon(x_0,x_1)=((1-\epsilon)f(x_0)+\epsilon
f(x_1),(1-\epsilon)f(x_1)+\epsilon f(x_0)),
\label{CML}
\end{equation}
for all $(x_0,x_1)\in\R^2$. In this paper, we consider local maps
defined by $f(x)=ax+(1-a)H(x-1/2)$ for all $x\in\R$ where $a>2$ and
where $H$ is the Heaviside function, i.e.\
\[
H(x)=\left\{\begin{array}{ccl}
0&\text{if}&x<0\\
1&\text{if}&x\geq 0
\end{array}\right.
\]
see Figure \ref{LOCAL_FIG}. 
The coupling parameter $\epsilon\in [0,1/2]$.

The dynamical system in $\R$ generated by the map $f$ has the
following properties. The set
$K=\bigcap\limits_{t\in\N}f^{-t}([0,1])$ is an invariant Cantor set
which contains all points with bounded orbit under $f$. The set $K$ is
called the repeller of the dynamical system $(\R,f)$. Moreover, the
dynamical system $(K,f)$ is topologically conjugated to 
the full shift on symbolic sequences of 0's and 1's. The coding map
associates to each point $x\in K$ a sequence of 0's and 1's using the
position of the iterates $f^t(x)$ with respect to $1/2$. (Conjugated
means the existence of a bijection between two systems and 
topologically conjugated means that this bijection is a
homeomorphism.)

These results can be obtained using nested pre-images of $[0,1]$ by
$f$ to construct each point of $K$. But since the map is piecewise
affine, one can use another method. The method consists in
solving the recurrence induced by $f$ to obtain a formal expression of
points in $\R$ which depends on symbolic sequences. This expression is
employed to deduce a condition for a symbolic sequence to code a point
in $K$ and to prove conjugacy to the symbolic system. Finally, the
analysis of these condition and expression show that the symbolic
system is the full shift and that $K$ is a Cantor set. (This method
and the results are contained in the analysis 
of the CML below when considering orbits lying in the
diagonal $\{(x,x)\ :\ x\in\R\}$. Indeed choosing $x_0=x_1$ in relation
(\ref{CML}) shows that the dynamics in the diagonal is given by $f$.)

Because of the definition of $F_\epsilon$, one can apply this second
method to the coupled map lattice to obtain a symbolic description of 
its repeller. In the next section, we present the first steps which
consist in obtaining the expression of points from their code, the
admissibility condition and in proving the conjugacy between the
dynamics in the repeller and the symbolic system. 

\subsection{Symbolic description of the coupled map lattice}
To express points in the repeller of the coupled map lattice using
symbolic sequences, we need to introduce some notations. Given a point 
$(x_0,x_1)\in\R^2$, let $\{{\bf x}^t\}_{t\in\N}$ be the orbit issued from
this point. It means that ${\bf x}^0=(x_0,x_1)$, ${\bf x}^t=(x_0^t,x_1^t)$ and
${\bf x}^{t+1}=F_\epsilon({\bf x}^t)$ for all
$t\in\N=\{0,1,2,\cdots\}$. Let also 
$\theta=\{\theta^t\}_{t\in\N}=\{\theta_0^t\theta_1^t\}_{t\in\N}\in\Omega$,
where $\Omega=\{00,01,10,11\}^\N$, be the corresponding code given by 
\begin{equation}
\theta^t_s=H(x_s^t-1/2),
\label{CODING}
\end{equation}
for all $t\in\N$ and $s\in\{0,1\}$. (We have chosen the notation
$\theta_0^t\theta_1^t$ rather than $(\theta_0^t,\theta_1^t)$ for the
sake of simplicity in the sequel.) The set $\Omega$ is endowed with
the product topology of the discrete topology on $\{00,01,10,11\}$.

Consider the norm $\|(x_0,x_1)\|=\max\{|x_0|,|x_1|\}$ in $\R^2$. The
{\bf repeller} of the CML is the set 
${\mathcal K}_\epsilon$ of points $(x_0,x_1)\in\R^2$ with bounded orbit,
i.e.\ for which $\sup\limits_{t\in\N}\|{\bf x}^t\|<+\infty$.

Let now $\epsilon_a=(a-1)/(2a)>1/4$
and, for every $\epsilon\in [0,\epsilon_a)$, consider the map
$\psi_\epsilon:\Omega\to\R$ defined by
\[
\psi_\epsilon(\theta)=(a-1)\sum_{k=0}^{+\infty}a^{-(k+1)}
\left(\ell_0^{(k)}\theta_0^k+\ell_1^{(k)}\theta_1^k\right),
\]
where $\ell_0^{(k)}=(1+(1-2\epsilon)^{-k})/2$ and
$\ell_1^{(k)}=(1-(1-2\epsilon)^{-k})/2$ for all $k\in\N$. The condition
$\epsilon\in [0,\epsilon_a)$ ensures that this series converges for
every $\theta\in\Omega$. (A computation of the derivative $DF_\epsilon$
of $F_\epsilon$ shows that the condition $\epsilon\in [0,\epsilon_a)$
is equivalent to the condition that both eigenvalues of $DF_\epsilon$
have modulus larger than 1 and hence to the condition that the mapping
$F_\epsilon$ is expanding.) {\bf In the rest of the paper, we assume
that $\epsilon\in [0,\epsilon_a)$.}

Finally, we introduce the shift $\sigma$ and the spatial flip $R$
acting in $\Omega$. Given a symbolic sequence $\theta$, let
$(\sigma\theta)^t=\theta^{t+1}$ and $(R\theta)^t=\theta_1^t\theta_0^t$
for every $t\in\N$. Note that $\sigma R=R\sigma$.

The map $\psi_\epsilon$ is used to obtain the expression of points in
${\mathcal K}_\epsilon$ and a condition for a symbolic sequence to code
for a point in this set. Given $\epsilon\in [0,\epsilon_a)$, let ${\mathcal
A}_\epsilon\subset\Omega$ be the set of sequences $\theta$ which
satisfy the following {\bf admissibility condition}
\begin{equation}
\theta_s^t=H(\psi_\epsilon(\sigma^tR^s\theta)-1/2),
\label{CONDADMI}
\end{equation}
for all $t\in\N$ and $s\in\{0,1\}$.  

With all these notations provided, we can state the conjugacy between
the symbolic systems and the CML restricted to its repeller. Given
$\theta\in\Omega$, let
$\phi_\epsilon(\theta)=(\psi_\epsilon(\theta),\psi_\epsilon(R\theta))$.
\begin{Pro} 
The map $\phi_\epsilon$ is a uniformly continuous bijection from ${\mathcal
A}_\epsilon$ to ${\mathcal K}_\epsilon$ and the following relation holds 
\[
F_\epsilon\circ\phi_\epsilon\big{|}_{{\mathcal A}_\epsilon}
=\phi_\epsilon\circ\sigma
\big{|}_{{\mathcal A}_\epsilon}.
\]
\label{CONJU}
\end{Pro}
\noindent
Despite this result, the topological entropies of $({\mathcal
A}_\epsilon,\sigma)$ and $({\mathcal K}_\epsilon,F_\epsilon)$ may differ
if ${\mathcal A}_\epsilon$ is not compact and this can happen as condition
(\ref{CONDADMI}) suggests. (For a recent discussion and references on
the topological entropy on non-compact sets, we refer to \cite{S01}.)
\smallskip

\noindent
{\sl Proof of the proposition:} The fact that $\phi_\epsilon$ is
uniformly continuous is an immediate consequence of uniform convergence in
the series defining $\psi_\epsilon$. 

We now prove that, for every point $(x_0,x_1)\in {\mathcal K}_\epsilon$,
there exists a sequence $\theta\in {\mathcal A}_\epsilon$ such that
$(x_0,x_1)=\phi_\epsilon(\theta)$.

Given $(x_0,x_1)\in\R^2$, let $L_\epsilon
(x_0,x_1)=((1-\epsilon)x_0+\epsilon x_1,(1-\epsilon)x_1+\epsilon
x_0)$. This operator $L_\epsilon$ is invertible when $\epsilon<1/2$
and $\|L_\epsilon^{-1}\|=(1-2\epsilon)^{-1}$. 

Let $(x_0,x_1)\in {\mathcal K}_\epsilon$ and let $\theta$ be the symbolic
sequence obtained by applying relation (\ref{CODING}). The
definition of $F_\epsilon$ then implies that for all $t\in\N$, we have
(when identifying $\theta^t=\theta_0^t\theta_1^t$ with
$(\theta_0^t,\theta_1^t)$ so that $L_\epsilon$ is well-defined on such
symbols)
\begin{equation}
{\bf x}^{t+1}=aL_\epsilon({\bf x}^t)+(1-a)L_\epsilon(\theta^t).
\label{ITER}
\end{equation}
We assume that $\epsilon<\epsilon_a$. Since $\epsilon_a<1/2$, we
can invert this relation to obtain 
\[
x_s^t=a^{-1}(L_\epsilon^{-1} {\bf x}^{t+1})_s+(a-1)a^{-1}\theta_s^t,
\] 
for all $t\in\N$ and $s\in\{0,1\}$. Iterating, we have for every
$t\in\N$ 
\begin{equation}
x_s^t=a^{-n}(L_\epsilon^{-n}{\bf x}^{t+n})_s+
(a-1)\sum_{k=0}^{n-1}a^{-(k+1)}(L_\epsilon^{-k}\theta^{t+k})_s,
\label{ITER2}
\end{equation}
for all $n\geq 1$ and $s\in\{0,1\}$. 
Using the norm of $L_\epsilon^{-1}$, it follows that 
\[
|a^{-n}(L_\epsilon^{-n}{\bf x}^{t+n})_s|\leq
(a(1-2\epsilon))^{-n}\|{\bf x}^{t+n}\|,
\]
where $s\in\{0,1\}$ and then
$\lim\limits_{n\to+\infty}a^{-n}(L_\epsilon^{-n}{\bf x}^{t+n})_s=0$
for every $(x_0,x_1)\in {\mathcal K}_\epsilon$ if $a(1-2\epsilon)>1$
i.e. if $\epsilon<\epsilon_a$.
In addition, one can check that 
$(L_\epsilon^{-k}{\bf x})_s=\ell_0^{(k)}x_s+\ell_1^{(k)}x_{1-s}$. It
results that if the point $(x_0,x_1)$ belongs to ${\mathcal K}_\epsilon$,
then by taking the limit $n\to+\infty$ in relation (\ref{ITER2}), we have
for all $t\in\N$ and $s\in\{0,1\}$
\[
x_s^t=(a-1)\sum_{k=0}^{+\infty}a^{-(k+1)}(L_\epsilon^{-k}\theta^{t+k})_s=
\psi_\epsilon(\sigma^tR^s\theta).
\]
In particular $(x_0,x_1)=\phi_\epsilon(\theta)$ and $\theta\in {\mathcal
A}_\epsilon$. Moreover, this relation shows that
$F_\epsilon(x_0,x_1)={\bf x}^1=\phi_\epsilon(\sigma\theta)$, i.e.\
the conjugacy holds. 

Let now $\theta\in {\mathcal A}_\epsilon$ and define ${\bf
x}(t)=\phi_\epsilon(\sigma^t\theta)$ for all $t\in\N$. By definition of
$\phi_\epsilon$, these points satisfy relation (\ref{ITER}). These
points also satisfy relation (\ref{CODING}) because $\theta\in
{\mathcal A}_\epsilon$. It results that we have
$F_\epsilon ({\bf x}(t))={\bf x}(t+1)$ for every $t$. Moreover,
the map $\phi_\epsilon$ is continuous and the set $\Omega$ is
compact. Therefore the sequence $\{{\bf x}(t)\}$ is a bounded orbit of
the CML and thus, ${\bf x}(0)\in {\mathcal K}_\epsilon$.

Furthermore, if $\phi_\epsilon(\theta_1)=\phi_\epsilon(\theta_2)$ where
$\theta_1,\theta_2\in {\mathcal A}_\epsilon$, then by relation
(\ref{CONDADMI}), we have $\theta_1=\theta_2$. Consequently $\phi_\epsilon$
is a bijection from ${\mathcal A}_\epsilon$ to ${\mathcal K}_\epsilon$. \qed

\subsection{Notations and properties of the conjugacy map}
We have shown that the CML dynamics in the repeller ${\mathcal
K}_\epsilon$ is entirely determined by ${\mathcal A}_\epsilon$. To
investigate this set, we need more notations and some properties of
the function $\psi_\epsilon$ which are collected in this section.

As usual, concatenations of symbols are called words and words can also
be concatenated. Given $\theta\in\Omega$ and $i\leq j\in\N$, the notation
$(\theta)_i^j=(\theta^i\theta^{i+1}\cdots\theta^j)$ denotes the word
composed of symbols from $i$ to $j$.

Superscripts are used to shorten the notation of a
$n$-fold concatenation of a symbol or a word. Parenthesis are
employed to separate symbols in a word or in a sequence. For
instance, $\theta=(01)(10)^n\Delta$ means
$\theta^0=01$, $\theta^t=10$ when $t\in\{1,\cdots,n\}$
and $\theta^t=\Delta^{t-n-1}$ when $t>n$ and $\theta=(01)^\infty$
means $\theta^t=10$ for all $t$. 

Moreover, we endow $\Omega$ with the following partial order. Firstly
comparing symbols,
we say that $\theta_0\theta_1\leq\Delta_0\Delta_1$ if $\theta_0=0$ and
$\Delta_0=1$ or if $\theta_0=\Delta_0$, $\theta_1=1$ and
$\Delta_1=0$ (in short terms $01\leq 00\leq 11\leq 10$). Now,
given two symbolic sequences $\theta,\Delta\in\Omega$, we say that
$\theta\leq\Delta$ if $\theta^t\leq \Delta^t$ for all $t\in\N$.
\begin{Pro}
(i) If $\theta\leq\Delta$, then
$\psi_\epsilon(\theta)\leq\psi_\epsilon(\Delta)$.

\noindent
(ii) There exists $k_\epsilon\geq 1$ such that if $\theta^t=\Delta^t$
for all $t\neq k,k+1$ for some $k\geq 1$ and if
\[
\left\{\begin{array}{lcl}
(\theta)_k^{k+1}=((10)(11))\quad \text{and}\quad
(\Delta)_k^{k+1}=((11)(10))&\text{if}&k\in\{1,\cdots,k_\epsilon\}\\
(\theta)_k^{k+1}=((11)(10))\quad \text{and}\quad
(\Delta)_k^{k+1}=((10)(11))&\text{if}&k>k_\epsilon
\end{array}\right.
\]
then $\psi_\epsilon(\theta)\leq\psi_\epsilon(\Delta)$.
\label{PROFI}
\end{Pro}
\noindent
Similarly as in \cite{R78}, the order on symbolic sequences will allow to
determine which sequences are admissible and which are not. But,
contrary to as for Lorenz maps, our ordering is only partial and one cannot
expect to characterise entirely the set of admissible sequences but
rather to obtain approximations or estimates. 
\smallskip

\noindent
{\sl Proof:} To prove (i), we have to show that for all $k\in\N$, we have
\[
\eta_k(01)\leq\eta_k(00)\leq\eta_k(11)\leq\eta_k(10),
\]
where
$\eta_k(\theta_0\theta_1)=\ell_0^{(k)}\theta_0+\ell_1^{(k)}\theta_1$.
We have $\eta_k(01)\leq\eta_k(00)$ and $\eta_k(11)\leq\eta_k(10)$
because $\ell_1^{(k)}\leq 0$ as its definition shows. In addition
$\eta_k(00)=0<1=\ell_0^{(k)}+\ell_1^{(k)}=\eta_k(11)$ and the
desired inequality follows.

Property (ii) is equivalent to the following one
\[
\left\{\begin{array}{lcl}
\mu_k((10)(11))\leq\mu_k((11)(10))&\text{if}&k\in\{1,\cdots,k_\epsilon\}\\
\mu_k((11)(10))\leq\mu_k((10)(11))&\text{if}&k> k_\epsilon
\end{array}\right.
\]
where $\mu_k(\theta^0\theta^1)=
\sum\limits_{i=k}^{k+1}a^{-i}\left(\ell_0^{(i)}\theta_0^{i-k}+
\ell_1^{(i)}\theta_1^{i-k}\right)$. Explicit calculations show that
\[
\mu_k((11)(10))-\mu_k((10)(11))=a^{-k}d(k),
\]
where $d(k)$ is a decreasing function of $k$, $d(1)\geq 0$ and
$d(+\infty)=-\infty$. Property (ii) then immediately follows. \qed

\section{Dynamics of piecewise affine expanding CML}
In this section, results of the analysis of the set of admissible
sequences corresponding to the CML (\ref{CML}) when $\epsilon$ varies
in $[0,\epsilon_a)$, together with the 
consequences for the CML are presented.

\subsection{Uncoupled domain}
Firstly, because of properties of $f$, the complete structure of the
uncoupled system is preserved for small coupling \cite{AF00}. That is
to say, we have ${\mathcal A}_\epsilon=\Omega$ when $\epsilon$ is small
enough. This property holds for more general maps $f$ than those
considered here. But the advantage of dealing with piecewise affine
maps is that the largest value of $\epsilon$ for which ${\mathcal
A}_\epsilon=\Omega$ can be computed.
\begin{Pro}
Let $\iota_a=(a-2)/(2a)>0$. Every sequence in $\Omega$ is admissible
iff $\epsilon\in [0,\iota_a)$.
\label{ALL}
\end{Pro}
\noindent
By compactness of $\Omega$, together with Proposition \ref{CONJU},
this result implies that the systems $({\mathcal
K}_\epsilon,F_\epsilon)$ and $({\mathcal A}_\epsilon,\sigma)$ are
topologically conjugated (and hence have the same topological entropy)
for every $\epsilon\in [0,\iota_a)$. The domain of couplings where
every sequence is admissible was also determined in \cite{C99} for
piecewise affine CML with infinite number of sites.
\smallskip

\noindent
{\sl Proof:} Let $\Omega_{(0)}\subset\Omega$ (resp.\
$\Omega_{(1)}\subset\Omega$) be the subset of sequences $\theta$ such
that $\theta^0_0=0$ (resp.\ $\theta^0_0=1$). It follows from the
admissibility condition that when the following inequalities
\[
\sup_{\theta\in\Omega_{(0)}}\psi_\epsilon(\theta)<1/2
\quad\text{and}\quad
\inf_{\theta\in\Omega_{(1)}}\psi_\epsilon(\theta)\geq 1/2,
\]
hold, every sequence in $\Omega$ is admissible. By continuity
of $\psi_\epsilon$ and by compactness of $\Omega$, these bounds are
minimum and maximum and thus, these inequalities are also
necessary conditions for the admissibility of every symbolic
sequence. In addition, the symmetry
$\psi_\epsilon(1-\theta)=1-\psi_\epsilon(\theta)$ implies that they
are equivalent to
$\sup\limits_{\theta\in\Omega_{(0)}}\psi_\epsilon(\theta)<1/2$.

Proposition \ref{PROFI}-(i) and the fact that $\ell_1^{(0)}=0$ imply
$\sup\limits_{\theta\in\Omega_{(0)}}\psi_\epsilon(\theta)=
\psi_\epsilon((0\omega)(10)^\infty)$ where
$\omega\in\{0,1\}$. Moreover explicit calculations show that
\begin{equation}
\psi_\epsilon((0\omega)(10)^\infty)=(a-1)\sum_{k=1}^{+\infty}
a^{-(k+1)}\ell_0^{(k)}=\frac{a-1-a\epsilon}{a(a-1-2a\epsilon)}<1/2
\quad\text{iff}\quad \epsilon<\iota_a.
\label{CRITIC}
\end{equation}
\qed
\smallskip

Note that relation (\ref{CRITIC}) and the fact that $\ell^{(k)}_1<0$ for every
$k\geq 1$ imply that ${\mathcal A}_{\iota_a}$ consists of all sequence
$\theta\in\Omega$ excepted those $\theta$ for which
there exist $t\in\N$ and $s\in\{0,1\}$ so that
$\sigma^tR^s\theta=(0\omega)(10)^\infty$ for some $\omega\in\{0,1\}$.

\subsection{Route to synchronisation}
In the domain $\epsilon\in (\iota_a,\epsilon_a)$, we have obtained
coupling dependent approximations of the set ${\mathcal A}_\epsilon$
rather than a complete characterisation. The approximations are chosen
in a family of sets obtained by restricting the length of words
$(10)^k$ and $(01)^k$ but by letting words of $00$'s and $11$'s be of
arbitrary length. 

\begin{figure}
\epsfxsize=5truecm
\centerline{\epsfbox{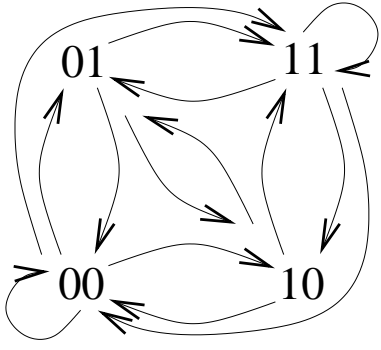}}
\caption{Associated graph to $\Lambda_1$.}
\label{GRAPHE}
\end{figure}

Precisely, the approximations belong to the family $\{\Omega_n\}$ defined as
follows. Firstly, consider the set $\Lambda_n$ of all sequences in
$\Omega$ for which the length of every word $(10)^k$ and $(01)^k$ does
not exceed $n$. (The graph associated to $\Lambda_1$ is represented
Figure \ref{GRAPHE}.)
Now, a sequence $\theta\in\Omega$
belongs to $\Omega_n$ if $\theta=(10)^\infty$, if $\theta=(01)^\infty$
or if there exist $k\in\N$ and $\Delta\in\Lambda_n$ such that either
$\theta=(01)^k\Delta$ or $\theta=(10)^k\Delta$.

The expression $\psi_\epsilon((0\omega)(10)^\infty)$
in relation (\ref{CRITIC}), together with continuity of the map
$\psi_\epsilon$ implies that, for every $\epsilon>\iota_a$, we have
$\psi_\epsilon((0\omega)(10)^n\theta)>1/2$ for every $\theta\in\Omega$
and every $\omega\in\{0,1\}$ if $n$ is sufficiently large. Therefore,
every sequence containing the word $(0\omega)(10)^{n+1}$, or by
symmetry the word $(1\omega)(01)^{n+1}$ is not admissible if $n$ is
sufficiently large. On the opposite, one can see that if a sequence
$\theta\in\Omega_n$ (where $n$ is any integer) contains the word
$(0\omega)(10)^k$ or $(1\omega)(01)^k$, then $k\leq n$.

The next statement describes the set of admissible sequences in the
domain $\epsilon\in (\iota_a,\epsilon_a)$. 
\begin{Thm}
There exists a right continuous decreasing function
$\underline{n}:(\iota_a,\epsilon_a)\to\N$ such that
$\lim\limits_{\epsilon\to\iota_a}\underline{n}(\epsilon)=+\infty$ and
$\lim\limits_{\epsilon\to\epsilon_a}\underline{n}(\epsilon)=0$ and
there exists a left continuous decreasing function
$\overline{n}:(\iota_a,\epsilon_a)\to\N$ such that
$\lim\limits_{\epsilon\to\iota_a}\overline{n}(\epsilon)=+\infty$ and
$\lim\limits_{\epsilon\to\epsilon_a}\underline{n}(\epsilon)=2$, so
that for every $\epsilon\in (\iota_a,\epsilon_a)$, we have
\[
\Omega_{\underline{n}(\epsilon)}\subsetneq {\mathcal A}_\epsilon\subset
\Omega_{\overline{n}(\epsilon)}.
\]
\label{DESC}
\end{Thm}
\begin{figure}[t]
\epsfxsize=11truecm
\centerline{\epsfbox{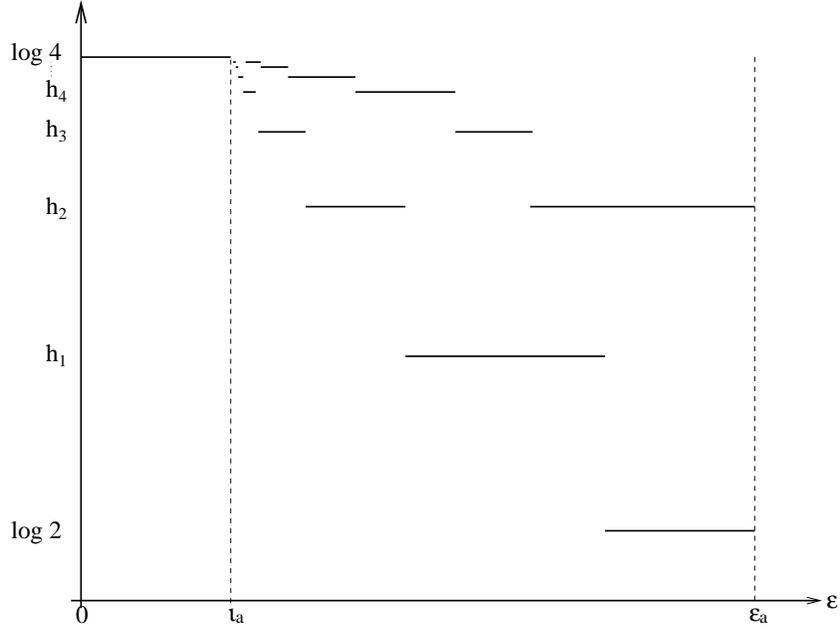}}
\caption{Upper bound $h_{\overline{n}(\epsilon)}$ and lower bound
$h_{\underline{n}(\epsilon)}$ for the topological entropy
$H_\epsilon$ of the CML versus coupling.} 
\label{BORNE}
\end{figure}
\noindent 
In other words, the dynamics becomes less chaotic and more homogeneous
when the coupling increases and this phenomenon is quantified by the
bounds $\Omega_{\underline{n}(\epsilon)}$ and 
$\Omega_{\overline{n}(\epsilon)}$. The decay of $\underline{n}$ and of
$\overline{n}$ indicate that the higher the coupling is, the shorter
the number of 
consecutive time steps in the region with symbol $10$ and the number
of consecutive time steps in the region $01$ have to be (excepted when
the initial condition belongs to such a region). It means that, when
the coupling increases, the orbits spend more time in regions where
both coordinates lie on the same side of 1/2, showing that the CML
gets closer to the synchronised regime. 

\noindent
(The strict inclusion $\Omega_{\underline{n}(\epsilon)}\subsetneq {\mathcal
A}_\epsilon$ implies that
$\underline{n}(\epsilon)<\overline{n}(\epsilon)$. However, we believe
in the existence of $M\in\N$ such that
$\overline{n}(\epsilon)\leq\underline{n}(\epsilon)+M$ for every 
$\epsilon$. One can actually prove the existence of such $M$ for every $a$
such that $\log a\geq 1$.)

A consequence of Theorem \ref{DESC} is the existence of upper bound and
lower bound for the topological entropy of the CML which are
decreasing functions of the coupling. Namely, if
$H_\epsilon=h({\mathcal K}_\epsilon,F_\epsilon)$ is the topological
entropy of the CML and if $h_n$ is the topological entropy of the
dynamical system $(\Omega_n,\sigma)$, then for all $\epsilon\in
(\iota_a,\epsilon_a)$, we have
\[
h_{\underline{n}(\epsilon)}\leq H_\epsilon\leq
h_{\overline{n}(\epsilon)}.
\]
Indeed, compactness of $\Omega_n$ implies that
$\phi(\Omega_{\underline{n}(\epsilon)})$ is
compact and then $\phi^{-1}_\epsilon$ is uniformly
continuous on this set. Since the topological entropy is invariant by
topological conjugacy provided the underlying spaces are compact (see
for instance \cite{KH95}) it results that $h_{\underline{n}(\epsilon)}= 
h(\phi_\epsilon(\Omega_{\underline{n}(\epsilon)}),F_\epsilon)$
and the left inequality follows. The right inequality is
an immediate consequence of conjugacy and uniform continuity of
$\phi_\epsilon$. 

In Appendix \ref{P-ENTROPY}, we show that the sequence $\{h_n\}$ has
the following properties: $h_{n+1}>h_n$ for every $n\in\N$ (and thus 
$h_{\underline{n}(\epsilon)}<h_{\overline{n}(\epsilon)}$), $h_0=\log 2$ and
$\lim\limits_{n\to+\infty}h_n=\log 4$. Together with Proposition
\ref{ALL}, these properties allow to construct the graphs
of $\epsilon\mapsto h_{\underline{n}(\epsilon)}$ and of
$\epsilon\mapsto h_{\overline{n}(\epsilon)}$, see Figure \ref{BORNE}.

A complementary result to Theorem \ref{DESC} is monotonicity of the
admissibility of symbolic sequences with coupling. The next statement
tells us that this assertion holds in a neighbourhood of $\iota_a$.
\begin{Pro}
There exists $\eta_a>\iota_a$ such that for every
$\epsilon_1<\epsilon_2\in [0,\eta_a)$, we have ${\mathcal
A}_{\epsilon_1}\supset {\mathcal A}_{\epsilon_2}$.
\label{MONOTONY}
\end{Pro}

The proofs of results in this section are organised as follows. In the
next section the existence of the function $\underline{n}$ is shown. The
existence of $\overline{n}$ is proved in Section \ref{P-NONADMIS}. The
proof of Proposition \ref{MONOTONY} is given in Appendix \ref{P-MONOTONY}.

\section{Existence of the function $\underline{n}$}
The explicit expression of $\psi_\epsilon$ and its properties allow to
show that, for every $n\in\N$, there exists a critical coupling
below which every sequence in $\Omega_n$ satisfies the admissibility
condition. Precisely, we have the following statement.
\begin{Pro}
There exists a strictly decreasing sequence
$\{\lambda_{a,n}\}_{n\in\N}$ with $\lambda_{a,0}=\epsilon_a$ and
$\lim\limits_{n\to+\infty}\lambda_{a,n}=\iota_a$ such
that, for every $n\in\N$, we have $\Omega_n\subset {\mathcal A}_\epsilon$
iff $\epsilon<\lambda_{a,n}$.
\label{ADMIS}
\end{Pro}
\noindent
The existence of the lower bound $\underline{n}$ in Theorem \ref{DESC}
as desired is a immediate consequence of this statement. Namely, we
have for every $\epsilon\in (\iota_a,\epsilon_a)$
\[
\underline{n}(\epsilon)=\max\{n\in\N\ :\ \epsilon<\lambda_{a,n}\},
\]
and this function is decreasing, right continuous and has range
$\N$. The proof that the inclusion
$\Omega_{\underline{n}(\epsilon)}\subset {\mathcal A}_\epsilon$ is strict
is given in section \ref{P-STRICT}.

Proposition \ref{ADMIS} is proved using the same arguments as those in the
proof of Proposition \ref{ALL}. Firstly, using properties of
$\psi_\epsilon$ and since each set $\Omega_n$ is invariant under
$\sigma$, under $R$ and under the change $\theta\mapsto 1-\theta$, one
proves that every sequence in $\Omega_n$ is admissible iff
$\sup\limits_{\theta\in\Omega_n\ :\
\theta^0_0=0}\psi_\epsilon(\theta)<1/2$. Secondly, the partial
order on $\Omega$ allows to prove that
the previous supremum is reached for the maximal sequence in
$\Omega_n$ with respect to the lexicographic order.
\begin{Lem}
For every $n\in\N$, we have
$\sup\limits_{\theta\in\Omega_n\ :\
\theta^0_0=0}\psi_\epsilon(\theta)=
\psi_\epsilon((00)((10)^n(11))^{\infty})$.
\label{MAXIMUM}
\end{Lem}
\noindent
Then, studying the behaviour of this maximum with coupling
gives the existence of the $\lambda_{a,n}$'s and their properties.
\begin{Lem}
There exists a strictly decreasing sequence
$\{\lambda_{a,n}\}_{n\in\N}$ with $\lambda_{a,0}=\epsilon_a$ and
$\lim\limits_{n\to+\infty}\lambda_{a,n}=\iota_a$ such
that, for every $n\in\N$, we have
$\psi_\epsilon((00)((10)^n(11))^{\infty})<1/2$ iff
$\epsilon<\lambda_{a,n}$.
\label{EXIS-LAM}
\end{Lem}
\noindent
The proofs of Lemmas \ref{MAXIMUM} and \ref{EXIS-LAM} are
given in sections below.

\subsection{Proof of Lemma \ref{MAXIMUM}}
By using the partial order, we reduce successively the set where to
find the supremum $\sup\limits_{\theta\in\Omega_n\ :\
\theta^0_0=0}\psi_\epsilon(\theta)$. 

Firstly, recall the subset
$\Lambda_n\subset\Omega_n$ of symbolic sequences for which the length
of every word $(10)^k$ and $(01)^k$ is not larger than $n$. Let
$\theta\in\Omega_n\setminus\Lambda_n$ be such that
$\theta_0^0=0$. Consider the sequence $\bar{\theta}$ obtained by
changing every symbol $01$ in $\theta$ by $00$. The sequence
$\bar{\theta}\in\Lambda_n$ and by Proposition \ref{PROFI}-(i), we have
$\psi_\epsilon(\theta)\leq\psi_\epsilon(\bar{\theta})$. These
arguments allow to show that 
\[
\sup\limits_{\theta\in\Omega_n\ :\
\theta^0_0=0}\psi_\epsilon(\theta)=\sup\limits_{\theta\in
\Lambda_n\:\ \theta_0^0=0}\psi_\epsilon(\theta).
\]

Since $\Lambda_0=\{00,11\}^\N$, the result of Lemma \ref{MAXIMUM}
immediately follows from Proposition \ref{PROFI}-(i) in the case
where $n=0$, i.e.\ $\sup\limits_{\theta\in
\Lambda_0\ :\
\theta_0^0=0}\psi_\epsilon(\theta)=\psi_\epsilon((00)(11)^\infty)$.
Therefore, in the rest of the proof, we assume that $n\geq 1$.

In a second step, we restrict to a set of sequences whose symbols,
for positive times, are either $10$ or $11$. This set $P_n$ is the
subset of sequences $\theta\in\Lambda_n$ with $\theta_0^0=0$ such
that, for all $t\geq 1$, $\theta^t\in\{11,10\}$ and if $\theta^t=11$,
there exists $i\in\{0,\cdots,\min\{n,t-1\}\}$ so that
$(\theta)_{t-i}^{t-i+n}=(10)^i(11)(10)^{n-i}$. Note that changing a
symbol $11$ by $10$ in an element of $P_n$ creates a sequence which
does not belong to $\Lambda_n$. Thus, $P_n$ is the smallest subset of
$\Lambda_n$ which can be obtained by only using Proposition \ref{PROFI}-(i)
when restricting and we have the following statement.
\begin{Lem}
For every $n\geq 1$, we have
$\sup\limits_{\theta\in\Lambda_n\ :\ \theta_0^0=0}
\psi_\epsilon(\theta)=\sup\limits_{\theta\in
P_n}\psi_\epsilon(\theta)$.
\label{LEM1}
\end{Lem}
\noindent
{\sl Proof:} As before, we show that for every $\theta\in\Lambda_n$
such that $\theta_0^0=0$, there exists $\bar{\theta}\in P_n$ such that
$\psi_\epsilon(\theta)\leq\psi_\epsilon(\bar{\theta})$. 

If $\theta\in\Lambda_n$ and $\theta_0^0=0$ but $\theta\not\in P_n$,
then there exists $t\geq 1$ such that $\theta^t\ne 10$ and
$(\theta)_{t-i}^{t-i+n}\ne (10)^i(11)(10)^{n-i}$ for all
$i\in\{0,\ldots,\min\{n,t-1\}\}$. Let $t(\theta)$ be the smallest of
such integers. 

Replace the symbol $\theta^{t(\theta)}$ by $10$ if this change does not
create a word $(10)^{n+1}$ in the resulting sequence, by $11$
otherwise. By Proposition \ref{PROFI}-(i), the resulting sequence, say
$A(\theta)$ (belongs to $\Lambda_n$ and) is such that
$\psi_\epsilon(\theta)\leq \psi_\epsilon(A(\theta))$ and
$t(A(\theta))>t(\theta)$. 

By iterating, we conclude that, for every $\theta\in\Lambda_n$ with
$\theta_0^0=0$, we have $\psi_\epsilon(\theta)\leq
\psi_\epsilon(\bar{\theta})$ where
$\bar{\theta}:=\lim\limits_{k\to+\infty}A^k(\theta)$ (the limit exists
because $t(A(\theta))>t(\theta)$). Because $t(A(\theta))>t(\theta)$,
we have $t(\bar{\theta})=+\infty$ and hence $\bar{\theta}\in P_n$. \qed
\smallskip

To restrict further the set on which the supremum is reached,
we now use Proposition \ref{PROFI}-(ii) in a similar reasoning. The
resulting set is the finite set $Q_n$ defined by 
\[
Q_n=\{(0\omega)(10)^i((11)(10)^n)^\infty\ :\ i\in\{0,\cdots,n\},\
\omega\in\{0,1\}\}
\]
\begin{Lem}
For every $n\geq 1$, we have
$\sup\limits_{\theta\in P_n}\psi_\epsilon(\theta)=\max\limits_{\theta\in
Q_n}\psi_\epsilon(\theta)$.
\label{APP-B}
\end{Lem}
\noindent
The proof is given in Appendix \ref{P-APP-B}. 
Finally, we show that the maximum is reached for the maximum in the
lexicographic order. 
\begin{Lem}
For every $n\geq 1$, we have $\max\limits_{\theta\in Q_n}
\psi_\epsilon(\theta)=\psi_\epsilon((00)((10)^n(11))^{\infty})$.
\end{Lem}
\noindent
{\sl Proof:} We show that, for every $i\in\{0,\cdots,n-1\}$, we have
\[
\psi_\epsilon((00)(10)^i((11)(10)^n)^{\infty})\leq
\psi_\epsilon((00)((10)^n(11))^{\infty}).
\]
By explicit calculations, one obtains that these inequalities are equivalent to
the following ones 
\[
\sum_{p=0}^{+\infty} a^{-(i+1+p(n+1))}\ell_1^{((i+1+p(n+1)))}\leq
\sum_{p=0}^{+\infty} a^{-(p+1)(n+1)}\ell_1^{((p+1)(n+1))},
\]
where $i\in\{0,\cdots,n-1\}$. Using the definition of $\ell_1^{(k)}$,
these in turn simplify to 
\begin{equation}
\sum_{k=0}^{n-i-1}a^k\sum_{j=n-i}^nb_\epsilon^j\leq\sum_{k=0}^{n-i-1}
b_\epsilon^k\sum_{j=n-i}^na^j,\quad i\in\{0,\cdots,n-1\},
\label{DERNI}
\end{equation}
where $b_\epsilon=a(1-2\epsilon)$. Since $1<b_\epsilon\leq a$ when $\epsilon\in
[0,\epsilon_a)$, for every $j\in\{n-i,\cdots,n\}$ the inequality
$\sum\limits_{k=0}^{n-i-1}a^{k-j}\leq\sum\limits_{k=0}^{n-i-1}b_\epsilon^{k-j}$
holds, i.e.\ we have $\sum\limits_{k=0}^{n-i-1}a^kb_\epsilon^j\leq 
\sum\limits_{k=0}^{n-i-1}b_\epsilon^ka^j$ which implies (\ref{DERNI}). \qed

\subsection{Proof of Lemma \ref{EXIS-LAM}}
The quantity $\psi_\epsilon((00)(11)^\infty)<1/2$ because $a>2$. In
other words, every sequence in $\Omega_0$ is admissible for every
$\epsilon<\epsilon_a$ and $\lambda_{a,0}=\epsilon_a$.

Let now $n\geq 1$ be fixed. We show that the map
$\epsilon\mapsto\psi_\epsilon((00)((10)^n(11))^{\infty})$ is strictly
increasing and diverges at $\epsilon_a$. Together with the fact that
\[
\psi_{\iota_a}((00)((10)^n(11))^{\infty})<
\psi_{\iota_a}((00)(10)^\infty)=1/2,
\]
this implies the existence of $\lambda_{a,n}\in (\iota_a,\epsilon_a)$.

We have
\[
\psi_\epsilon((00)((10)^n(11))^{\infty})=\frac{a-1}{2a}\left(\frac{1}{a-1}+
\frac{1}{b_\epsilon-1}+\frac{1}{a^{n+1}-1}
-\frac{1}{b_\epsilon^{n+1}-1}\right),
\]
where again $b_\epsilon=a(1-2\epsilon)>1$. Hence
$\lim\limits_{\epsilon\to\epsilon_a}
\psi_\epsilon((00)((10)^n(11))^{\infty})=+\infty$ because $b_\epsilon\to 1$
when $\epsilon\to\epsilon_a$. Moreover, explicit calculations shows
that the sign of the derivative $\psi'_\epsilon((00)((10)^n(11))^{\infty})$
is the same as the sign of
\[
(b_\epsilon^{n+1}-1)^2-(n+1)b_\epsilon^n(b_\epsilon-1)^2=
(b_\epsilon-1)^2\left(\left(\sum_{k=0}^nb_\epsilon^k\right)^2-
(n+1)b_\epsilon^n\right),
\]
and the latter is positive because $b_\epsilon>0$. 

To prove the inequality $\lambda_{a,n+1}<\lambda_{a,n}$, $n\in\N$, it
suffices to show that
\[
\psi_\epsilon((00)((10)^n(11))^\infty)<
\psi_\epsilon((00)((10)^{n+1}(11))^\infty).
\]
Using the expression of
$\psi_\epsilon((00)((10)^n(11))^\infty)$, the latter is equivalent to
\[
\frac{1}{b_\epsilon^{n+1}-1}-\frac{1}{b_\epsilon^n-1}<
\frac{1}{a^{n+1}-1}-\frac{1}{a^n-1},
\]
where $n\geq 1$, and hence equivalent to the property that the
the map $a\mapsto (a^{n+1}-1)^{-1}-(a^n-1)^{-1}$ is strictly
increasing when $a>1$. Computing its derivative, we obtain that this
condition holds iff
$n(\sum\limits_{k=0}^na^k)^2>(n+1)a(\sum\limits_{k=0}^{n-1}a^k)^2$,
which holds for every $a>1$ and $n\geq 1$.

Each $\lambda_{a,n}$ can be defined as $\sup\{\epsilon\ :\
\psi_\epsilon((00)((10)^n(11))^\infty)<1/2\}$. The sequence
$\{\lambda_{a,n}\}$ is decreasing and bounded below. Therefore it
converges to a limit which, by continuity of
$\psi_\epsilon$, is equal to $\sup\{\epsilon\ :\
\psi_\epsilon((00)(10)^\infty)<1/2\}=\iota_a$.

\subsection{Proof of the strict inclusion
$\Omega_{\underline{n}(\epsilon)}\subsetneq 
{\mathcal A}_\epsilon$}\label{P-STRICT}
To that goal, we prove the existence, for every $n\geq 2$, of a sequence in the
complement set of $\Omega_n$ which belongs to ${\mathcal A}_\epsilon$ for every
$\epsilon<\lambda_{a,n}$ where $\lambda_{a,n}$ is defined in
Lemma \ref{EXIS-LAM}. The desired sequence is the periodic
sequence $((01)^{n+1}(10)^{n+1})^\infty$.

Indeed by applying the maps $\sigma$ and $\theta\mapsto 1-\theta$, a
necessary and sufficient condition for
$((01)^{n+1}(10)^{n+1})^\infty\in {\mathcal A}_\epsilon$ is
\[
\max\limits_{p\in\{1,\cdots,n+1\}}
\psi_\epsilon((01)^p((10)^{n+1}(01)^{n+1})^\infty)<1/2.
\]
Recall now the function $\mu_k$ employed in the proof of
Proposition \ref{PROFI}-(ii). It can be shown that the condition
$\mu_k((01)(10))<\mu_k((10)(01))$ for all $k\in\N$ holds for every
$\epsilon<\epsilon_a$. Using this property repeatedly, we obtain
\begin{eqnarray*}
\psi_\epsilon((01)^p((10)^{n+1}(01)^{n+1})^\infty)&<&
\psi_\epsilon((01)^{p-1}((10)(01)(10)^n(01)^n)^\infty)\\
&<&\psi_\epsilon((01)^{p-1}((10)^2(01)(10)^{n-1}(01)^n)^\infty)\\
&<&\cdots\\
&<&\psi_\epsilon((01)^{p-1}((10)^{n+1}(01)^{n+1})^\infty)
\end{eqnarray*}
showing that the previous maximum is reached for $p=1$. Moreover, using the
property $\mu_k((10)(01))<\mu_k((11)(10))$ for all $k\in\N$, which holds
for every $\epsilon\in (\iota_a,\epsilon_a)$, a similar argument
shows that
\[
\psi_\epsilon((01)((10)^{n+1}(01)^{n+1})^\infty)<
\psi_\epsilon((01)((10)^n(11)(10)(01)^n)^\infty).
\]
Since $(01)((10)^n(11)(10)(01)^n)^\infty\in\Omega_n$, this inequality
implies that 
\[\psi_\epsilon((01)((10)^{n+1}(01)^{n+1})^\infty)<1/2,
\]
for every $\epsilon\in (\iota_a,\lambda_{a,n})$ and thus
$\Omega_{\underline{n}(\epsilon)}\subsetneq {\mathcal A}_\epsilon$.

\section{Existence of the function $\overline{n}$}\label{P-NONADMIS}
As for the lower bound, to obtain the upper bound
$\overline{n}$ in Theorem \ref{DESC}, we prove the existence of values
of the coupling above which every sequence in the
complement set of $\Omega_n$ does not satisfy the admissibility
condition. The main statement of this section is the following one.
\begin{Pro}
There exists a decreasing sequence $\{v_{a,n}\}_{n\geq 2}$ with
$v_{a,n}<\epsilon_a$ and $\lim\limits_{n\to+\infty}v_{a,n}=\iota_a$
such that, for every $n\geq 2$, we have ${\mathcal A}_\epsilon\subset\Omega_n$
if $\epsilon\in (v_{a,n},\epsilon_a)$.
\label{NONADMIS}
\end{Pro}
\noindent
Similarly as for the lower bound, Proposition \ref{NONADMIS} implies
the existence of the function $\overline{n}$ with the desired
properties. This function is defined by
\[
\overline{n}(\epsilon)=\min\{n\geq 2\ :\ v_{a,n}<\epsilon\},
\]
for every $\epsilon\in (\iota_a,\epsilon_a)$.

The conclusion of Proposition \ref{NONADMIS} does not hold for $n=0$, i.e.\
we have $v_{a,0}=\epsilon_a$. This is because explicit calculations show that
the sequence $((01)(10))^\infty$ is admissible for every $\epsilon\in
[0,\epsilon_a)$. We do not know if this statement holds for $n=1$.

Let $V_n$ be the complement set of $\Omega_n$ in $\Omega$. The proof
of Proposition \ref{NONADMIS} uses the following sufficient condition
for every sequence in $V_n$ not to belong to ${\mathcal A}_\epsilon$: for
every $\theta\in\Omega$, either we have
$\psi_\epsilon((0\omega)(10)^{n+1}\theta)\geq 1/2$ or we have
$\psi_\epsilon((10)\theta)<1/2$.

(Note that the following sufficient condition
\[
\inf\limits_{\theta\in\Omega}\psi_\epsilon((00)(10)^{n+1}\theta)=
\psi_\epsilon((00)(10)^{n+1}(01)^\infty)\geq 1/2,
\]
does not hold when $\epsilon$ is close to $\epsilon_a$ because
$\lim\limits_{\epsilon\to\epsilon_a}\psi_\epsilon((00)(10)^{n+1}(01)^\infty)=
-\infty$. Therefore, it cannot be used to prove the existence of
$v_{a,n}$. The same comment applies to the improved condition
\[
\psi_\epsilon((00)(10)^{n+1}((01)^{n+1}(00))^\infty)\geq 1/2.)
\]

To employ this sufficient condition, we need some
notations and an auxiliary result. Let the functions
$S(\theta)=(2a)^{-1}(a-1)\sum\limits_{k=0}^{+\infty}
a^{-k}(\theta_0^k+\theta_1^k)$ and
$D_\epsilon(\theta)=(2a)^{-1}(a-1)\sum\limits_{k=0}^{+\infty}
b_\epsilon^{-k}(\theta_0^k-\theta_1^k)$ which is well-defined when
$\epsilon<\epsilon_a$. Recall that $b_\epsilon=a(1-2\epsilon)$
decreases in $(1,2)$ when $\epsilon$ increases in
$(\iota_a,\epsilon_a)$.  

Using the definitions of $\psi_\epsilon$ and of $\ell_0^{(k)}$ and
$\ell_1^{(k)}$ we obtain for every $\omega\in\{0,1\}$ 
\[
\psi_\epsilon((0\omega)(10)^{n+1}\theta)=(2a)^{-1}(a-1)\sum_{k=1}^{n+1}(a^{-k}
+b_\epsilon^{-k})+a^{-(n+2)}S(\theta)+b_\epsilon^{-(n+2)}D_\epsilon(\theta),
\]
and $\psi_\epsilon((10)\theta)=
a^{-1}(a-1)+a^{-1}S(\theta)+b_\epsilon^{-1}D_\epsilon(\theta)$.
Consider also the positive quantity
$M_{a,n}=a^{-1}(a-1)+2^{n+1}a^{-(n+2)}$ and the following statement.
\begin{Lem}
For every $n\geq 1$, there exists $\rho_{a,n}<\epsilon_a$ such that
for every $\epsilon\in (\rho_{a,n},\epsilon_a)$ and every
$\theta\in\Omega$ so that $D_\epsilon(\theta)\in [-1/2,M_{a,n}]$, we have
$\psi_\epsilon((10)\theta)<\psi_\epsilon((0\omega)(10)^{n+1}\theta)$.
\label{INEQ}
\end{Lem}
\noindent
{\sl Proof of Lemma \ref{INEQ}:} Using the notations above, one shows that
the inequality
$\psi_\epsilon((10)\theta)<\psi_\epsilon((0\omega)(10)^{n+1}\theta)$
is equivalent to the following one
\begin{eqnarray*}
\lefteqn{b_\epsilon^{-1}(1-b_\epsilon^{-(n+1)})D_\epsilon(\theta)
+(2a)^{-1}(a-1)(n+1-\sum_{k=1}^{n+1}b_\epsilon^{-k})}\\
& &<(2a)^{-1}(a-1)(\sum_{k=1}^{n+1}a^{-k}+n-1)-a^{-1}(1-a^{-(n+1)})
S(\theta).
\end{eqnarray*}
The left hand side equals 0 when $b=1$. Using the inequality
$S(\theta)\leq 1$, one proves that for every $a>2$ and $n\geq 2$, the
following inequality holds (and hence the previous inequality holds
for $\epsilon=\epsilon_a$ if $|D_{\epsilon_a}(\theta)|<+\infty$)
\[
(2a)^{-1}(a-1)(\sum_{k=1}^{n+1}a^{-k}+n-1)-a^{-1}(1-a^{-(n+1)})
S(\theta)>0.
\]
Since the assumption on $\theta$ forces
$|D_\epsilon(\theta)|\leq\max\{1/2,|M_{a,n}|\}$ and since this bound
does not depend on $\epsilon$, by continuity with $\epsilon$, there exists
$\rho_{a,n}<\epsilon_a$ and so that for every $\epsilon\in
(\rho_{a,n},\epsilon_a)$, we have
$\psi_\epsilon((10)\theta)<\psi_\epsilon((0\omega)(10)^{n+1}\theta)$ 
for every $\theta\in\Omega$ such that $D_\epsilon(\theta)\in
[-1/2,M_{a,n}]$. \qed 
\smallskip

\noindent
{\sl Proof of Proposition \ref{NONADMIS}:} Let
$\bar{v}_{a,n}=\max\{\iota_a,\rho_{a,n}\}$ and assume that $\epsilon\in
(\bar{v}_{a,n},\epsilon_a)$. Given any sequence
$\Delta\in V_n$, there exists $(s,t)$ such that the
sequence $\sigma^tR^s\Delta$ either equals $(0\omega)(10)^{n+1}\theta$ or
equals $(1\omega)(01)^{n+1}\theta$ for some $\theta\in\Omega$ and
$\omega\in\{0,1\}$. Assume that
$\sigma^tR^s\Delta=(0\omega)(10)^{n+1}\theta$, the other case can be
completed by using the symmetry
$\psi_\epsilon(1-\theta)=1-\psi_\epsilon(\theta)$.
One of the following conditions holds.
\begin{itemize}
\item[(a)] $D_\epsilon(\theta)>M_{a,n}$,
\item[(b)] $D_\epsilon(\theta)\in [-1/2,M_{a,n}]$,
\item[(c)] $D_\epsilon(\theta)<-1/2$.
\end{itemize}
In case (a), using the inequalities $S(\theta)\geq 0$ and
$b_\epsilon<2$ in the expression of $\psi_\epsilon((0\omega)(10)^{n+1}\theta)$
and using the definition of $M_{a,n}$, we obtain for every $\epsilon>\iota_a$,
\[
\psi_\epsilon((0\omega)(10)^{n+1}\theta)>
(2a)^{-1}(a-1)\sum_{k=1}^{n+1}(a^{-k}+2^{-k})+
2^{-(n+2)}M_{a,n}=1/2,
\]
showing that $\Delta$ does not belong to ${\mathcal
A}_\epsilon$. Similarly, in case (c), since $S(\theta)\leq 1$, using
the expression of $\psi_\epsilon((10)\theta)$, we obtain
\[
\psi_\epsilon((10)\theta)\leq 1+D_\epsilon(\theta)<1/2,
\]
which gives the same conclusion. Finally in case (b), one can apply
Lemma \ref{INEQ} to conclude that either
$\psi_\epsilon((0\omega)(10)^{n+1}\theta)\geq 1/2$ or
$\psi_\epsilon((10)\theta)<1/2$. In all cases, we have shown that the original
sequence $\Delta$ does not belong to ${\mathcal A}_\epsilon$.

We have shown that no sequence in $V_n$ belongs to ${\mathcal A}_\epsilon$
when $\epsilon\in (\bar{v}_{a,n},\epsilon_a)$. Let $v_{a,n}$ be the
infimum of couplings $\epsilon$ such that no sequence in $V_n$ belongs
to ${\mathcal A}_\epsilon$ for all $\epsilon\in (v_{a,n},\epsilon_a)$.
By Proposition \ref{ALL}, we have $v_{a,n}\geq\iota_a$ for every
$n\geq 2$. Moreover, since $V_{n+1}\subset V_n$, we have $v_{a,n+1}\leq
v_{a,n}$ and thus, the limit $v_a:=\lim\limits_{n\to+\infty}v_{a,n}$
exists. Finally, continuity of $\psi_\epsilon$ implies that $v_a=\iota_a$.
\qed

\section{Dynamics of coupled piecewise increasing maps lattices}
As already said, the existence of a conjugacy between the weakly
coupled system and the uncoupled system has been proved for more general
CML than only piecewise affine ones. In this section, we show that
the results of Theorem \ref{DESC} also extend to more
general CML.

These CML are defined as in (\ref{CML}) but instead of the
piecewise affine map, we consider any map $f:\R\to\R$ which is
continuous and increasing on $(-\infty,1/2)$ and on $[1/2,+\infty)$
and which is such that there exists $a_f>1$ so that for all
$x,y\in\R$, we have 
\begin{equation}
|f(x)-f(y)|\geq a_f|x-y|.
\label{EXPAN}
\end{equation}
In the sequel, the notation $a_f$ will always means the largest of
such numbers and $F_{f,\epsilon}$ will denote the CML defined using
such a map $f$.

\subsection{Symbolic description}
As for the piecewise affine system, the first step of analysis of
$F_{f,\epsilon}$ is to
show that, due to expansiveness (\ref{EXPAN}), the system $({\mathcal
K}_{f,\epsilon},F_{f,\epsilon})$ (where ${\mathcal K}_{f,\epsilon}$ still
denote the repeller of the CML) admits a symbolic description. Let
$\epsilon_f=(a_f-1)/(2a_f)>0$.
\begin{Pro}
For every $\epsilon\in [0,\epsilon_f)$, there exists a map
$\psi_{f,\epsilon}:\Omega\to\R$ such that ${\mathcal
K}_{f,\epsilon}=\phi_{f,\epsilon}({\mathcal A}_{f,\epsilon})$ where
$\phi_{f,\epsilon}(\theta)=
(\psi_{f,\epsilon}(\theta),\psi_{f,\epsilon}(R\theta))$
and where ${\mathcal A}_{f,\epsilon}$ is the set of sequences $\theta$ which
satisfy the admissibility condition
\[
\theta_s^t=H(\psi_{f,\epsilon}(\sigma^tR^s\theta)-1/2),
\]
for all $t\in\N$ and $s\in\{0,1\}$. Moreover, the map
$\phi_{f,\epsilon}$ is uniformly continuous, one-to-one and conjugates the
systems $({\mathcal K}_{f,\epsilon},F_{f,\epsilon})$ and $({\mathcal
A}_{f,\epsilon},\sigma)$.
\end{Pro}
\noindent
{\sl Proof:} Let $\epsilon\in [0,\epsilon_f)$ and let
$(x_0,x_1)$ be a point in ${\mathcal K}_{f,\epsilon}$. Let $f_0$
(resp.\ $f_1$) be the restriction of $f$ to
$(-\infty,1/2)$ (resp.\ $[1/2,+\infty)$). The inequality
$\epsilon_f<1/2$ ensures that $L_\epsilon$ is invertible. The
definition of $F_{f,\epsilon}$ then shows that the following relation
holds between the iterates of the orbit $\{{\bf x}^t\}$ issued from $(x_0,x_1)$
\[
x_s^t=(G_{\theta^t}({\bf x}^{t+1}))_s,
\]
for all $t\in\N$ and $s\in\{0,1\}$. Here
$\theta^t=\theta^t_0\theta^t_1$ where $\theta_s^t=H(x_s^t-1/2)$ are
the components of the code $\theta$ and the map $G_{\theta^0}$ is defined by
\begin{equation}
G_{\theta^0}({\bf x})=(f^{-1}_{\theta_0^0}((L_\epsilon^{-1}{\bf x})_0),
f^{-1}_{\theta_1^0}((L_\epsilon^{-1}{\bf x})_1)).
\label{DEFIG}
\end{equation}
By iterating it results that
\[
x_s^t=(G_{\theta^t}\circ
G_{\theta^{t+1}}\circ\cdots\circ G_{\theta^{t+n}}({\bf x}^{t+n+1}))_s.
\]
Let us prove that these components only depend on $\sigma^t\theta$,
i.e.\ on $\{\theta^n\}_{n\geq t}$. Firstly, given $t$, the sequence
$\{G_{\theta^t}\circ
G_{\theta^{t+1}}\circ\cdots\circ G_{\theta^{t+n}}({\bf x}^{t+n+1})\}_{n\in\N}$
is constant, so its limit exists. Moreover, relation (\ref{EXPAN}) and
$\|L_\epsilon^{-1}\|=(1-2\epsilon)^{-1}$ imply the inequality
\begin{equation}
\|G_{\theta^0}({\bf x})-G_{\theta^0}({\bf y})\|\leq
(a_f(1-2\epsilon))^{-1}\|{\bf x}-{\bf y}\|.
\label{CONTRAC}
\end{equation}
Since $(a_f(1-2\epsilon))^{-1}<1$ when $\epsilon<\epsilon_f$, it
results that
\[
\lim_{n\to+\infty}\|G_{\theta^t}\circ
G_{\theta^{t+1}}\circ\cdots\circ G_{\theta^{t+n}}({\bf x}^{t+n+1})-
G_{\theta^t}\circ G_{\theta^{t+1}}\circ\cdots\circ
G_{\theta^{t+n}}({\bf y}^{t+n+1})\|=0,
\]
if the orbits $\{{\bf x}^t\}$ and $\{{\bf y}^t\}$ are bounded
and have the same code. Consequently, the limit only depends on the sequence
$\sigma^t\theta$, that is to say
$x_s^t=\tilde\psi(\sigma^t\theta,s)$. Moreover, the relation
(\ref{DEFIG}) shows that
$\tilde\psi(R\theta,s)=\tilde\psi(\theta,1-s)$. Denoting by
$\psi_{f,\epsilon}(\theta)$ the quantity $\tilde\psi(\theta,0)$, we
conclude that every point in $(x_0,x_1)\in {\mathcal K}_{f,\epsilon}$ writes
$(x_0,x_1)=\phi_{f,\epsilon}(\theta)$, that the sequence $\theta\in
{\mathcal A}_{f,\epsilon}$ and that
$F_{f,\epsilon}(\phi_{f,\epsilon}(\theta))=\phi_{f,\epsilon}(\sigma\theta)$.

Now, to prove that every $\theta\in {\mathcal A}_{f,\epsilon}$ codes a
point in ${\mathcal K}_{f,\epsilon}$, we
first linearly extend the monotonic components of $f$ over the whole
$\R$, i.e.\ we set $f_0(x)=a_f(x-1/2)+f_0(1/2-0)$ for $x\geq
1/2$ and $f_1(x)=a_f(x-1/2)+f_1(1/2)$ if $x<1/2$. This
allows to define the image $\psi_{f,\epsilon}(\theta)$ for every
$\theta\in\Omega$ as the following limit
\[
\psi_{f,\epsilon}(\theta):=\lim_{t\to+\infty}(G_{\theta^0}\circ\cdots\circ
G_{\theta^t}({\bf x}))_0.
\]
Indeed the choice of extensions for the monotonic components of $f$
ensures that contraction (\ref{CONTRAC}) holds for every pair ${\bf
x},{\bf y}$ of points in $\R^2$ and every symbol. It results that if
it exists, the previous limit does not depend on ${\bf x}\in\R^2$.

To show that the limit exists, we note that the monotony and
expansiveness of the $f_s$'s imply the existence of $C>0$ such that
$|f_s^{-1}(x)|\leq a_f^{-1}|x|+C$ for every $x\in\R$. Consequently, we have
\[
\|G_{\theta^0}({\bf x})\|\leq
(a_f(1-2\epsilon))^{-1}\|{\bf x}\|+C(1-2\epsilon)^{-1},
\]
and then the sequence $\{G_{\theta^0}\circ\cdots\circ
G_{\theta^t}({\bf x})\}$ is bounded. This property, together with the
contraction (\ref{CONTRAC}), allows to prove that this sequence is a
Cauchy sequence and hence has a limit in $\R^2$.

Once $\psi_{f,\epsilon}(\theta)$ has been defined for every symbolic
sequence, one can develop similar arguments to those in the proof of
Proposition \ref{CONJU} to show that if
$\theta_s^t=H(\psi_{f,\epsilon}(\sigma^tR^s\theta)-1/2)$ for every
$s,t$, then $\{\phi_{f,\epsilon}(\sigma^t\theta)\}$ is a bounded orbit of
$F_{f,\epsilon}$. That is to say, we have $\phi_{f,\epsilon}({\mathcal
A}_{f,\epsilon})\subset {\mathcal K}_\epsilon$.

Finally, one easily proves that the map $\phi_{f,\epsilon}$ is
one-to-one and uniformly continuous and the proof is complete. \qed

\subsection{Approximations of symbolic systems}
We now show that, up to arbitrary shifts of the functions $\underline{n}$ and
$\overline{n}$, the approximations obtained for the piecewise
affine map also hold for the symbolic dynamics of CML
with local map sufficiently close to the piecewise affine one. This
means that small perturbations of the local map result in small
perturbations of (approximations of) the complete dynamical picture
when the coupling increases and then in bounds on the function
``topological entropy versus coupling''.

\begin{figure}[t]
\epsfxsize=11truecm
\centerline{\epsfbox{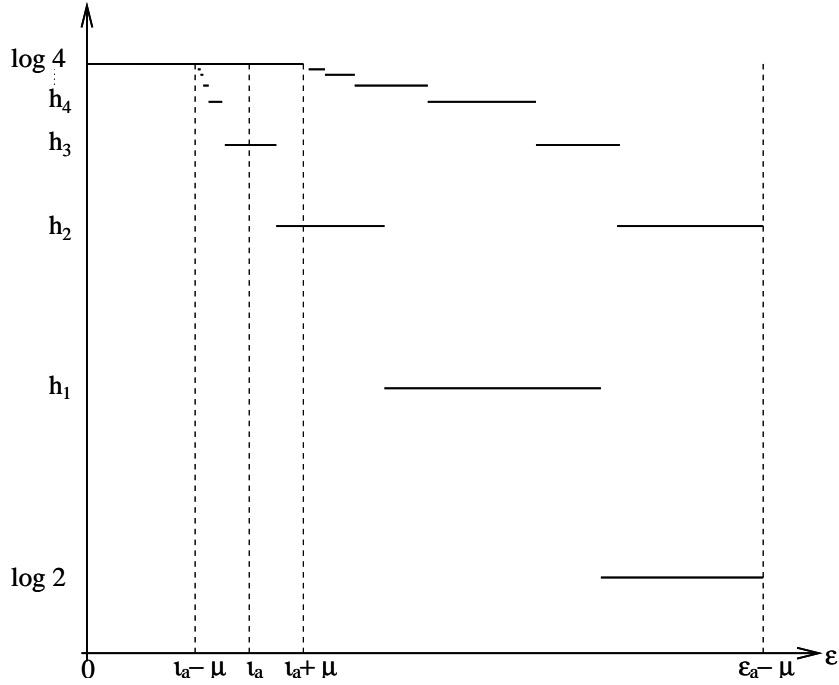}}
\caption{Upper bound and lower bound for the topological entropy
of $({\mathcal K}_{f,\epsilon},F_{f,\epsilon})$ versus coupling when $f$
is such that $a_f\in (a_\mu,a)$ and $\|f-f_a\|<\delta$ (see Theorem
\ref{LAST}).} 
\label{BORNE2}
\end{figure}

We shall denote by $f_a$ the piecewise affine map $x\mapsto
ax+(1-a)H(x)$. The existence of approximations for more general CML
and their continuous dependence with the local map in the
$C^1$-topology is given in the following statement.
\begin{Thm}
For every $\mu>0$ sufficiently small, there exists $a_\mu\in (1,a)$
such that for every $\alpha\in (a_\mu,a)$, there exists $\delta>0$ so that
every piecewise increasing expanding map $f$ with $a_f=\alpha$
and $\|f-f_a\|<\delta$ is such that $\epsilon_f>\epsilon_a-\mu$ and
satisfies the following properties:  

\noindent
(i) ${\mathcal A}_{f,\epsilon}=\Omega$ if $\epsilon\in
[0,\iota_a-\mu)$ and ${\mathcal A}_{f,\epsilon}\subsetneq\Omega$ if
$\epsilon\in (\iota_a+\mu,\epsilon_a-\mu)$,

\noindent
(ii) $\Omega_{\underline{n}^\mu(\epsilon)}\subsetneq {\mathcal
A}_{f,\epsilon}$ for every $\epsilon\in (\iota_a-\mu,\epsilon_a-\mu)$
where $\underline{n}^\mu$ is a right continuous decreasing function
with range $\N$ satisfying $\underline{n}^\mu(\epsilon)\geq
\underline{n}(\epsilon+\mu)$ for every $\epsilon\in
(\iota_a-\mu,\epsilon_a-\mu)$, 

\noindent
(iii) ${\mathcal
A}_{f,\epsilon}\subset\Omega_{\overline{n}^\mu(\epsilon)}$ for every 
$\epsilon\in (\iota_a+\mu,\epsilon_a-\mu)$ where $\overline{n}^\mu$ is
a left continuous decreasing function satisfying
$\overline{n}^\mu(\epsilon)\leq \overline{n}(\epsilon-\mu)$ for every
$\epsilon\in (\iota_a+\mu,\epsilon_a-\mu)$.
\label{LAST}
\end{Thm}
\noindent
As in the piecewise affine case, this result allow to obtain graphs of
upper and lower bounds for the topological entropy of $({\mathcal
K}_{f,\epsilon},F_{f,\epsilon})$, see Figure \ref{BORNE2}.

The existence of a partial order as in Proposition
\ref{PROFI} for general map $f$ (even when restricting the
domain of couplings) is unknown. Hence, instead of using order on
symbolic sequences, the proof of Theorem \ref{LAST} uses
continuity 
of the map $\psi_{f,\epsilon}(\theta)$ with $f$ uniformly in
$\theta$ and strengthening of the admissibility and non-admissibility
conditions of the sets $\Omega_n$. 
\smallskip

\noindent
{\sl Proof:} We denote by $\psi_{a,\epsilon}$, the map
$\psi_{f_a,\epsilon}$. Uniform continuity of $\psi_{f,\epsilon}$ is
given in the next statement.
\begin{Pro}
Let $f$ be such that $\|f-f_a\|<\delta$. Then ($a_f\leq a$ and) for every
$\epsilon\in [0,\epsilon_f)$ and every $\theta\in\Omega$, we have
\[
|\psi_{a,\epsilon}(\theta)-\psi_{f,\epsilon}(\theta)|\leq
\frac{\delta}{a(1-(a_f(1-2\epsilon))^{-1})}.
\]
\label{CONTI}
\end{Pro}
\noindent
{\sl Proof of the proposition:} Given a map $f$, let again
$f_0$ (resp.\ $f_1$) be the increasing function obtained by linearly
continuating the left (resp.\ right) monotonic component. The
assumption $\|f-f_a\|<\delta$ implies that the differences of
inverse maps are bounded as follows
\[
\max\left\{\|f_0^{-1}-(f_a)_0^{-1}\|,\|f_1^{-1}-(f_a)_1^{-1}\|\right\}<
\delta/a.
\]
Consequently, for every $x\in\R^2$, we have
\[
\|G_{\theta^0}^a({\bf x})-G_{\theta^0}^f({\bf x})\|\leq\delta/a,
\]
where $G_{\theta^0}^a$ means the map defined in relation (\ref{DEFIG})
and $G_{\theta^0}^f$ means the corresponding map obtained with $f$
instead of $f_a$. Using also relation (\ref{CONTRAC}) for
$G_{\theta^t}^f$, we obtain for all $t\in\N$
\begin{eqnarray*}
\lefteqn{\|G_{\theta^0}^a\circ\cdots\circ G_{\theta^t}^a({\bf x})-
G_{\theta^0}^f\circ\cdots\circ G_{\theta^t}^f({\bf x})\|}\\
& &\leq
\|G_{\theta^0}^a\circ G_{\theta^1}^a\circ\cdots\circ G_{\theta^t}^a({\bf x})-
G_{\theta^0}^f\circ G_{\theta^1}^a\circ\cdots\circ G_{\theta^t}^a({\bf
x})\|\\
& &+
\|G_{\theta^0}^f\circ G_{\theta^1}^a\circ\cdots\circ G_{\theta^t}^a({\bf x})
-G_{\theta^0}^f\circ G_{\theta^1}^f\circ\cdots\circ
G_{\theta^t}^f({\bf x})\|\\
& &\leq\delta/a+b_f^{-1}\|G_{\theta^1}^a\circ\cdots\circ
G_{\theta^t}^a({\bf x})-
G_{\theta^1}^f\circ\cdots\circ G_{\theta^t}^f({\bf x})\|
\end{eqnarray*}
where $b_f=a_f(1-2\epsilon)$. By induction, this inequality shows
that, for all $t\in\N$, we have
\[
\|G_{\theta^0}^a\circ\cdots\circ G_{\theta^t}^a({\bf x})-
G_{\theta^0}^f\circ\cdots\circ G_{\theta^t}^f({\bf x})\|
\leq\frac{\delta}{a}\sum_{k=0}^tb_f^{-k}.
\]
The desired conclusion then follows when taking the limit
$t\to+\infty$. \qed
\smallskip

We now present a proof of statements {\sl (i)} and {\sl (ii)} of the Theorem.

For every $n\in\N$, the map
$\psi_{a,\epsilon}((00)((10)^n(11))^\infty)$ is a strictly increasing
function of $\epsilon$. By definition of $\lambda_{a,n}$ (Lemma
\ref{EXIS-LAM}), it results that for every $\eta>0$ (smaller than
$1/2-1/a$), there exists $\lambda_{a,n}^\eta\in (0,\lambda_{a,n})$
such that for every $\epsilon\in [0,\lambda_{a,n}^\eta)$, we have
\[
\psi_{a,\epsilon}((00)((10)^n(11))^\infty)<1/2-\eta.
\]

The map $\psi_{a,\epsilon}((00)((10)^n(11))^\infty)$ is a strictly increasing
function of $n$ (proof of Lemma \ref{EXIS-LAM}). Thus,
$\lambda_{a,n+1}^\eta<\lambda_{a,n}^\eta$. By continuity, we conclude
that $\lim\limits_{n\to+\infty}\lambda_{a,n}^\eta=\iota_a^\eta$ where
$\iota_a^\eta$ is such that, for every $\epsilon\in [0,\iota_a^\eta)$,
we have $\psi_{a,\iota_a^\eta}((00)(10)^\infty)<1/2-\eta$.

In addition, the following limit exist $\lim\limits_{\eta\to
0}\lambda_{a,n}^\eta=\lambda_{a,n}$ for every $n$ and $\lim\limits_{\eta\to
0}\iota_a^\eta=\iota_a$.

All these limits and the limit
$\lim\limits_{n\to+\infty}\lambda_{a,n}=\iota_a$ can be used to show
that for every $\mu>0$, there exists $\eta_\mu$ such that
\begin{equation}
\sup\limits_{n\in\N}\lambda_{a,n}-\lambda_{a,n}^{\eta_\mu}<\mu.
\label{SUPREM}
\end{equation}

Given $\mu\in (0,\epsilon_a)$, let $a_\mu$ be defined by
$a_\mu(1-2\lambda_{a,0}^{\eta_\mu})=1$
and, given $\alpha\in (a_\mu,a)$, let $\delta>0$ be defined by
$\eta_\mu=\frac{\delta}{a(1-(\alpha(1-2\lambda_{a,0}^{\eta_\mu}))^{-1})}$.

By Proposition \ref{CONTI}, by definition of $\lambda_{a,n}^\eta$ and
by Lemma \ref{MAXIMUM}, it results that, for any map $f$ with $a_f=\alpha$ and
$\|f-f_a\|<\delta$ (by relation (\ref{SUPREM}), we have
$\epsilon_f>\lambda_{a,n}^{\eta_\mu}>\epsilon_a-\mu$) and for any $\epsilon\in
(0,\lambda_{a,n}^{\eta_\mu})$, we have $\psi_{f,\epsilon}(\theta)<1/2$
for all $\theta\in\Omega_n$ such that $\theta_0^0=0$. In short terms,
$\Omega_n\in {\mathcal A}_{f,\epsilon}$ for every $\epsilon\in
(0,\lambda_{a,n}^{\eta_\mu})$.

The inequality (\ref{SUPREM}) implies that
$\iota_a^{\eta_\mu}>\iota_a-\mu$ (we assume that $\mu<\iota_a$). By
definition of 
$\iota_a^{\eta_\mu}$, we conclude that $\psi_{f,\epsilon}(\theta)<1/2$
for every $\theta\in\Omega$ with $\theta_0^0=0$ if $\epsilon\in
(0,\iota_a-\mu)$. The first assertion of Statement {\sl (i)} is
proved. The second assertion follows from Statement {\sl (iii)} or can
be directly proved by using that $\eta_\mu$ can be chosen so that 
$\psi_{a,\epsilon}((00)(10)^\infty)>1/2+\eta_\mu$ for every
$\epsilon>\iota_a+\mu$.

When $\epsilon\in (\iota_a-\mu,
\lambda_{a,0}^{\eta_\mu})$, consider the quantity
$\underline{n}^\mu(\epsilon)=\sup\{n\in\N\ :\
\epsilon<\lambda_{a,n}^{\eta_\mu}\}$. We have 
$\Omega_{\underline{n}^\mu(\epsilon)}\subsetneq {\mathcal
A}_{f,\epsilon}$ (the strict inequality also comes from the
admissibility of $((01)^{\underline{n}^\mu(\epsilon)+1}
(10)^{\underline{n}^\mu(\epsilon)+1})^\infty$) and the inequality
(\ref{SUPREM}) 
implies that
$\underline{n}^\mu(\epsilon)\geq\underline{n}(\epsilon+\mu)$ for every
$\epsilon\in (\iota_a-\mu,\epsilon_a-\mu)$. Statement {\sl (ii)} is proved.

Statement {\sl (iii)} can be proved using the same strategy. That is
to say, one has to investigate properties of the smallest numbers
$v_{a,n}^\eta\geq v_{a,n}$ which are such that, for every $\epsilon\in
(v_{a,n}^\eta,\epsilon_a)$ and every $\theta\in\Omega$, either we have
$\psi_{a,\epsilon}((00)(10)^{n+1}\theta)<1/2-\eta$ or we have
$\psi_{a,\epsilon}((10)\theta)>1/2-\eta$. The proof is left to the
reader. \qed

\appendix
\section{Proof of properties of the sequence $\{h_n\}$}\label{P-ENTROPY}
Firstly, the quantity $h_n$ is equal to the topological entropy of
$(\Lambda_n,\sigma)$ because $\Lambda_n$ is the non-wandering set of
the system $(\Omega_n,\sigma)$ and $\sigma$ is a continuous map (see
e.g.\ \cite{R99}).

Since $(\Lambda_n,\sigma)$ is a subshift of finite type, $h_n$ is the
exponential rate of increase of $N_t$, the number of admissible words
in $\Lambda_n$ of length $t$, precisely
$h_n=\limsup\limits_{t\to+\infty}\frac{\log N_t}{t}$.

To obtain an equation for $h_n$ when $n\geq 1$ (the case where
$n=0$ can be computed directly), we compute induction relations for $N_t^0$
(resp. $N_t^{01}$), the number of admissible words in $\Lambda_n$ of
length $t$ ending with 00 or with 11 (resp. with 01 but not
with $(01)^2$). Let also $N_t^{(01)^k}$ (resp.\ $N_t^{(10)^k}$) be the number
of admissible words of length $t$ ending with $(01)^k$ but not with
$(01)^{k+1}$ (resp.\ with $(10)^k$ but not with $(10)^{k+1}$) where
$k\in\{1,\cdots,n\}$.

By symmetry, we have $N_t^{(01)^k}=N_t^{(10)^k}$ for any
$k\in\{1,\cdots,n\}$. In addition, it is easy to see that
$N_{t+1}^{(01)^k}=N_t^{(01)^{k-1}}$ for all $k\in\{2,\cdots,n\}$. Now,
since any word can be followed by 00 or 11, we have
\[
N_{t+1}^0=2N_t^0+4\sum_{k=1}^nN_t^{(01)^k}
=2N_t^0+4\sum_{k=0}^{n-1}N_{t-k}^{01}
\]
Similarly, we successively obtain
\[
N_{t+1}^{01}=N_t^0+\sum\limits_{k=0}^nN_t^{(10)^k}
=N_t^0+\sum\limits_{k=0}^nN_t^{(01)^k}
=N_t^0+\sum\limits_{k=0}^{n-1}N_{t-k}^{(01)}.
\]
These relations show that, if $N_t^0$ and $N_t^{01}$ increase
exponentially, then they rates are equal. Assuming that
$N_t^0=c_0\lambda^t$ and $N_t^{01}=c_1\lambda^t$ in these relations
forces $\lambda$ to be a solution of $f_n(\lambda)=0$ where
\[
f_n(\lambda)=\lambda^{2(n+1)}-2\lambda^{2n+1}-(\lambda+2)\sum_{k=1}^n
\lambda^{k+n}.
\]
In other words, we have $h_n=\log\lambda_n$ where $\lambda_n$
is the largest solution of $f_n(\lambda)=0$. The properties of the
sequence $\{h_n\}$ can now be proved.

Because the alphabet has 4 symbols, each $\lambda_n$ is at most 4. But
$f_n(4)>0$ and then $\lambda_n<4$ for all $n\geq 1$. Moreover
$f_n(2)<0$ and then $\lambda_n> 2$ for every $n\geq
1$. The case $n=0$ can be achieved by direct calculations which show
that $f_0(\lambda)=\lambda-2$ and then $\lambda_0=2$.

In addition the definition of $\lambda_n$ implies that
$f_{n+1}(\lambda_n)=\lambda_n^{2n+3}(\lambda_n-4)<0$ and then
$\lambda_{n+1}>\lambda_n$ since $f_{n+1}(\lambda)>0$ for $\lambda$
sufficiently large.

Finally, we have $\lim\limits_{n\to+\infty}f_n(\lambda)=-\infty$ for
every $\lambda\in (2,4)$ and thus $\lim\limits_{n\to+\infty}\lambda_n=4$.

\section{Proof of Proposition \ref{MONOTONY}}\label{P-MONOTONY}
We are going to prove the existence of $\eta_a>\iota_a$ such that, for
every sequence $\theta\in\Omega$ so that $\theta_0^0=0$, either the
derivative with respect to $\epsilon$, $\psi'_\epsilon(\theta)$ is
positive for every 
$\epsilon\in (\iota_a,\eta_a)$ or we have $\psi_\epsilon(\theta)<1/2$
for every $\epsilon\in (\iota_a,\eta_a)$. By symmetry
$\psi_\epsilon(1-\theta)=1-\psi_\epsilon(\theta)$, it results that
every sequence not
in ${\mathcal A}_{\epsilon_1}$ is not in ${\mathcal A}_{\epsilon_2}$  for
every $\epsilon_2>\epsilon_1$ and the Proposition follows. 

A direct computation shows that $\psi'_\epsilon(\theta)$ is a
continuous map of $\theta$ 
for every $\epsilon$. Using also that
$\psi'_{\iota_a}((0\omega)(10)^\infty)>0$ as follows from relation
(\ref{CRITIC}), there exists $t_a\in\N$ such that
for every $\Delta\in\Omega$, we have 
\[
\psi'_{\iota_a}((0\omega)(10)^{t_a}\Delta)\geq 
\frac{\psi'_{\iota_a}((0\omega)(10)^\infty)}{2}.
\] 
Moreover, it can be shown that the family
$\{\psi'_\epsilon(\theta)\}_{\theta\in\Omega}$ is
equi-continuous for every $\epsilon$. Together with the previous
inequality, we conclude that there exists $\eta_1>\iota_a$ such that
for every $\epsilon\in (\iota_a,\eta_1)$ and every $\Delta\in\Omega$, we have 
\[
\psi'_{\epsilon}((0\omega)(10)^{t_a}\Delta)>0.
\]
In addition, by Proposition \ref{PROFI}-(i), we have
$\psi_\epsilon(\theta)\leq\max\limits_{1\leq
i<t_a}\psi_\epsilon((0\omega)(10)^i(11)(10)^\infty)$ for every sequence
$\theta=(0\omega)(10)^k\Delta$ with $k<t_a$ and $\Delta^0\neq 10$. The
fact that $\ell_1^{(k)}<0$ for $k>0$ and relation (\ref{CRITIC})
then show that
\[
\max\limits_{1\leq i<t_a}\psi_{\iota_a}((0\omega)(10)^i(11)(10)^\infty)<1/2.
\]
By continuity of the map $\epsilon\mapsto\psi_\epsilon(\theta)$, there
exists $\eta_2>\iota_a$ such that for every $\epsilon\in
(\iota_a,\eta_2)$, we have $\max\limits_{1\leq
i<t_a}\psi_\epsilon((0\omega)(10)^i(11)(10)^\infty)< 1/2$. Letting
$\eta_a=\min\{\eta_1,\eta_2\}$ gives the desired statement. 

\section{Proof of Lemma \ref{APP-B}}\label{P-APP-B}
As in the proof of Lemma \ref{LEM1}, we introduce a mapping which
allows to prove that for every $\theta\in P_n$, there exists
$\bar{\theta}\in Q_n$ such that
$\psi_\epsilon(\theta)\leq\psi_\epsilon(\bar{\theta})$.

Given $\theta\in P_n$, consider the first occurrence when
the word 11 is not followed by $(10)^n$
\[
T(\theta)=\inf\left\{t\geq 1\ : \ \theta^t=11\ \text{and}\ \min
\left\{p>0\ :\ \theta^{t+p}=11\right\}\leq n\right\}.
\]
Clearly, we have $Q_n=\{\theta\in P_n\ :\
T(\theta)=+\infty\}$. If $T(\theta)<+\infty$, then let
\[
k(\theta)=\min \{k>T(\theta) \ :\ \theta^k=11\}.
\]
Note that the definition of $P_n$ imposes that
$T(\theta)>1$. (Otherwise, the sequence writes
$(0\omega)(11)(10)^k(11)\cdots$ for some $k\in\{0,\cdots n-1\}$
and then does not belong to $P_n$.) Similarly, the definition
of $T(\theta)$ imposes that $\theta^{k(\theta)+1}=10$. (Otherwise, the
definition of $P_n$ forces $\theta^{k(\theta)}$ to be preceeded by
$(10)^n$ which contradicts the definition of $T(\theta)$.)

To define the mapping, we use the following subsets
\[
P_\leq=\{\theta\in P_n\ :\ T(\theta)\leq k_\epsilon\}\quad\text{and}\quad
P_>=\{\theta\in P_n\ :\ k_\epsilon<T(\theta)< +\infty\},
\]
where $k_\epsilon$ is given in Proposition \ref{PROFI}-(ii). Let the
map $B$ defined in $P_n$ by $B(\theta)=\theta$ if $\theta\in Q_n$, by
\[
B(\theta)^t=\left\{
\begin{array}{lcl}
\theta^t&\text{if}&t\neq T(\theta)-1,T(\theta)\\
11&\text{if}&t=T(\theta)-1\\
10&\text{if}&t=T(\theta)
\end{array}\right.
\]
for all $t\in\N$ if $\theta\in P_\leq$ and by
\[
B(\theta)^t=\left\{
\begin{array}{lcl}
\theta^t&\text{if}&t\neq k(\theta),k(\theta)+1\\
10&\text{if}&t=k(\theta)\\
11&\text{if}&t=k(\theta)+1
\end{array}\right.
\]
for all $t\in\N$ if $\theta\in P_>$.

The map $B$ is such that
$\psi_\epsilon(\theta)\leq\psi_\epsilon(B(\theta))$ for every
$\theta\in P_n$.

We now detail the study of the behaviour of $T(\theta)$ under
iterations. Firstly,
note that every sequence $\theta\in P_\leq\cup P_>$ writes
\[
\theta=(0\omega)(10)^i((11)(10)^n)^j(11)(10)^k(11)(10)^l(11)\cdots,
\]
where $i\in\{0,\cdots,n\}$, $j\in\N$, $k\in\{0,\cdots,n-1\}$,
$l\in\{0,\cdots,n\}$, $i+k\geq n$ if $j=0$ and $k+l\geq n$. In
particular, the symbol $\theta^{T(\theta)}$ is the 11 following
$((11)(10)^n)^j$, the integer $k=k(\theta)-T(\theta)-1$ and the
inequalities $i+k\geq n$ if $j=0$ and $k+l\geq n$ ensure that
$\theta\in P_n$.

Assume that $\theta\in P_\leq$ and consider the case where
$j>0$. We have
\[
B(\theta)=(0\omega)(10)^i((11)(10)^n)^{j-1}(11)(10)^{n-1}(11)(10)^{k+1}
(11)(10)^l(11)\cdots,
\]
and by iterating
\[
B^j(\theta)=(0\omega)(10)^i(11)(10)^{n-1}((11)(10)^n)^{j-1}(11)(10)^{k+1}
(11)(10)^l(11)\cdots.
\]
Now if $i>0$, then one can apply $B$ once again to obtain
\[
B^{j+1}(\theta)=(0\omega)(10)^{i-1}((11)(10)^n)^j(11)(10)^{k+1}
(11)(10)^l(11)\cdots.
\]
If $i=0$, then $B^j(\theta)\in O_n\setminus P_n$ so we apply the map
$A$ defined in the proof of Lemma \ref{LEM1} to
obtain (recall that $T$ is only defined for sequences in $P_n$)
\[
(A\circ B^j)(\theta)=(0\omega)(10)^n((11)(10)^n)^{j-1}(11)(10)^{k+1}
(11)(10)^l(11)\cdots\in P_n
\]
In both cases, the resulting value of $T$ at least equals the original
one and, if $k+1=n$, is larger than the original one. If $k+1<n$, one
has to apply the same reasoning once again to conclude by iterating
that in the case where $\theta\in P_\leq$ and $j\geq 0$, if $i\geq
n-k$, then
\[
B^{(n-k)(j+1)}(\theta)=(0\omega)(10)^{i-(n-k)}
((11)(10)^n)^{j+1}(11)(10)^l(11)\cdots,
\]
which shows that $T(B^{(n-k)(j+1)}(\theta))\geq T(\theta)+k+1$. (The
case $j=0$, for which we certainly have $i\geq n-k$, can be treated
analogously.) If $i<n-k$, then
\[
(A\circ (B^{i(j+1)+j}))(\theta)=(0\omega)(10)^n((11)(10)^n)^{j-1}(11)
(10)^{k+i+1}(11)(10)^l(11)\cdots,
\]
and either $T((A\circ (B^{i(j+1)+j}))(\theta))\geq T(\theta)+k+1$ if
$k+i+1=n$ or $(A\circ (B^{i(j+1)+j}))(\theta)$ is a sequence satisfying
the assumptions of the previous case where $i\geq n-k$.

Assume now that $\theta\in P_>$. Then, we have (note that the
conditions $k+l\geq n$ and $k<n$ impose that $l>0$)
\[
B(\theta)=(0\omega)(10)^i((11)(10)^n)^j(11)(10)^{k+1}(11)(10)^{l-1}(11)(10)^m
(11)(10)^p(11)\cdots,
\]
where $m,p\in\{0,\cdots,n\}$, $l+m\geq n$ and $m+p\geq n$. Either $k+1=n$ and
$T(B(\theta))\geq T(\theta)+1$ or $k+1<n$. In the latter case, if $l-1+m\geq
n$, then one can apply $B$. Otherwise, together with the inequality $l+m\geq
n$, we have $l+m=n$ and then
\[
(A\circ B)(\theta)=(0\omega)(10)^i((11)(10)^n)^j(11)(10)^{k+1}(11)(10)^n
(11)(10)^p(11)\cdots,
\]
for which we apply the same analysis. Generally one proves
that if $\theta\in P_>$ and $k+l+m\geq 2n$ (i.e.\ $l-(n-k)+m\geq n$)
then
\[
B^{n-k}(\theta)=(0\omega)(10)^i((11)(10)^n)^{j+1}(11)(10)^{l-(n-k)}(11)(10)^m
(11)(10)^p\cdots,
\]
and $T(B^{n-k}(\theta))\geq T(\theta)+n+1$. If $k+l+m<2n$, then
\begin{eqnarray*}
\lefteqn{(A\circ B^{l+m-n+1})(\theta)=}\\
& &(0\omega)(10)^i((11)(10)^n)^j(11)
(10)^{k+l+m-n+1}(11)(10)^n(11)(10)^p(11)\cdots.
\end{eqnarray*}
If $k+l+m-n+1=n$, the desired result follows. If $k+l+m-n+1<n$, then
either $k+l+m-n+1+n+p\geq 2n$ and the desired result follows
by applying $B^{n-(k+n+m-n+1)}$ or $k+l+m+1+p<2n$ and
we have the same alternative as before. Since $k+l+m+p+1>k+l+m$, by
repeating the process, we are sure that the resulting sequence
satisfies the inequality $k+l+m\geq 2n$ after a finite number of
iterations.

We have shown that, for every sequence $\theta\in
P_\leq\cup P_>$, there exists an integer $k$ such that
\[
T((A\circ
B)^k(\theta))>T(\theta)\quad\text{and}\quad
\psi_\epsilon(\theta)\leq\psi_\epsilon((A\circ B)^k(\theta)).
\]
As in the proof of Lemma \ref{LEM1}, we conclude that the limit
$\lim\limits_{k\to+\infty}(A\circ B)^k(\theta)$ exists, belongs to
$Q_n$ and has a larger value of $\psi_\epsilon$. 

\section*{Acknowledgements}
We want to thank E.\ Ugalde for stimulating discussions.

\end{document}